# Localized Ecological Momentary Assessment for Mental Health Research in China: An Implementation-Oriented Framework and Preliminary Case Application

Xinying Zhao[1,2*] , BSc, MEd, PDEng; Yue Li[1*] , BE; Jiafeng Wang[1*] , BMed, MM; Yunfan Fu[1] , BEd, MEd; Ruilin Guo[3] , MIM; Chen Yang[3] , MSc; Cheng Yao[2] , PhD; Wei Deng[1,4] , MD

[1]Affiliated Mental Health Center & Hangzhou Seventh People's Hospital, School of Medicine, Zhejiang University, Hangzhou, Zhejiang, China
[2] School of Computer and Computing Science, Zhejiang University, Hangzhou, Zhejiang, China
[3]Hangzhou ZenSeven Technology Co., Ltd., Hangzhou, Zhejiang, China
[4]Liangzhu Laboratory, MOE Frontier Science Center for Brain Science and Brain-Machine Integration, State Key Laboratory of Brain-Machine Intelligence, Zhejiang University, Hangzhou, Zhejiang, China
* these authors contributed equally

## Abstract

**Background:** Ecological momentary assessment (EMA) is increasingly used in mental health research, but research-grade deployment requires platforms supporting protocol configuration, automated delivery, participant management, and data export. In China, these requirements are not consistently supported.

Objective: We aimed to identify workflow gaps affecting localized EMA deployment, develop an implementation-oriented framework for platform assessment, and assess Huixin EMAI.

**Methods:** We reviewed EMA platforms reported in Chinese mental health studies in CNKI and Wanfang. A multidisciplinary panel of 6 experts developed the Multi-dimensional EMA Platform Evaluation Framework (MEPEF) and benchmarked 7 platforms across 43 indicators in 6 domains. MEPEF was then applied to Huixin EMAI using deployment logs from 48 participants, questionnaires from 44 participants, and semistructured interviews with 6 researchers.

**Results:** We identified 66 empirical studies. Most relied on instant-messaging-based workflows (36/66, 54.5%), whereas specialized EMA platforms were less common (14/66, 21.2%). MEPEF provided a 6-domain framework for cross-platform benchmarking and highlighted a trade-off between localized deployability and advanced research functions. In a Huixin EMAI deployment, 1893 of 2472 expected prompts were completed (76.6%), with a median response latency of 4.0 minutes (Q1-Q3 0.0-13.0). Participant feedback indicated favorable acceptability; researchers reported support for core workflows but gaps in control, delivery monitoring, and data readiness.

**Conclusions:** The main challenge for EMA in Chinese mental health research appears to lie less in feasibility than in recurring workflow gaps affecting localized deployment. This study translates these gaps into structured evaluation and design targets, providing an implementation-oriented pathway for advancing localized EMA platforms.

## Introduction

### Background

Mental health symptoms are inherently dynamic and often fluctuate across daily contexts and routines. However, many clinical and research assessments still rely on retrospective self-report at infrequent time points, which is vulnerable to recall bias and may fail to capture within-day changes in symptoms, affect, and context [1]. Ecological momentary assessment (EMA) addresses this limitation by collecting repeated self-reports in real time and naturalistic settings, thereby improving ecological validity and enabling intensive longitudinal assessment of symptom trajectories [2,3].

EMA has become increasingly feasible through smartphone-based prompting, timestamped responses, and passive sensing, and its integration with wearable biosensors has further supported digital phenotyping by linking lived experience with physiological and contextual data streams [4–6]. These capabilities are particularly relevant to mental health research, where symptom expression is shaped by time-varying environmental, behavioral, and physiological factors. However, their effective use depends not only on the availability and compatibility of smartphones and wearable devices but also on EMA platform capabilities such as scheduling, adherence monitoring, participant management, data export, and support for passive or wearable data [7]. These capabilities are especially important in multimodal digital phenotyping, which requires consistent data capture, synchronization, processing, and feature construction across heterogeneous devices and deployment settings. Recent reviews have therefore emphasized the need for greater standardization in these workflows [8].

Despite the availability of established international EMA platforms, implementing research-grade EMA at scale in China remains challenging. International platforms may face barriers related to local deployment, accessibility, and data-hosting practices, whereas localized tools often provide limited support for advanced protocol configuration, automated monitoring, and multimodal integration with wearable devices [7,9–14]. As a result, EMA in Chinese mental health research may still depend on combinations of survey tools and messaging-based workflows that are locally accessible but often labor-intensive, relatively inflexible, and difficult to sustain across studies. These limitations affect not only observational EMA studies but also the longer-term potential to support more adaptive EMA-based applications, including ecological momentary interventions and just-in-time adaptive interventions (JITAIs), which require timely, context-sensitive data streams and reliable implementation workflows. Recent reviews suggest that JITAI-based mental health interventions show promise, but the field remains methodologically early and continues to face challenges in intervention logic, implementation, and evaluation [15].

Against this background, an important gap remains between knowing which EMA tools are available and understanding whether they can support real-world mental health research workflows in practice. In the Chinese context, evidence is still limited regarding the tools currently used, the workflow-level gaps they leave unresolved, and the conditions required for feasible localized deployment. These implementation frictions matter for mental health research because they can directly affect adherence, data completeness, symptom-timing precision, and the interpretability of intensive longitudinal evidence. Because these challenges extend across the EMA study workflow, they are not easily addressed through platform feature comparison alone. A workflow-aligned framework with relevant indicators is therefore needed to organize recurring gaps into structured evaluation domains, support cross-platform benchmarking, and inform design targets for localizing EMA platforms. Although this study focuses on Chinese mental health research, the underlying challenge is broader and applies to other contexts in which implementing EMA requires balancing research-grade functionality with local constraints, cross-device compatibility, and privacy-sensitive data handling.

### Objectives

This study aimed to identify workflow-level gaps affecting localized EMA deployment in Chinese mental health research, develop an implementation-oriented framework for structured platform assessment, and derive framework-guided design targets for localized EMA platforms, with Huixin EMAI serving as an existing localized case for assessing current implementation coverage and preliminary implementation evidence in China.

## Methods

### Study Overview

We first identified workflow-level gaps affecting localized EMA deployment in Chinese mental health research through a landscape review. These findings informed the development of an implementation-oriented evaluation framework, which was then used to benchmark existing platforms and characterize their strengths and limitations. Based on the landscape review and benchmarking findings, framework-guided design targets for localized EMA platforms were derived. The framework was subsequently applied to Huixin EMAI, an existing localized platform, to assess current implementation coverage, with the derived design targets used to map relevant platform functions and support a preliminary implementation evaluation using participant-side and researcher-side evidence.

## Landscape Review of EMA Platforms and Workflow-Level Implementation Gaps in Chinese Mental Health Research

We conducted an implementation-focused structured landscape review to identify EMA platforms and related implementation workflows reported in Chinese mental health research. Because the primary aim of this stage was to characterize tools and workflow practices used under local deployment conditions in China, we searched 2 major Chinese academic databases, China National Knowledge Infrastructure (CNKI) and Wanfang Data, from database inception to October 31, 2025. Searches used full-text terms for EMA methods ("生态瞬时评估" OR "经验取样法"), with topic restriction to mental health ("心理健康"). Detailed eligibility criteria are presented in Textbox 1.

**Textbox 1.** Eligibility Criteria for the Landscape Review

**Inclusion criteria**

- Empirical studies in psychiatry, psychology, or mental health research.
- Use of ecological momentary assessment, experience sampling method, or equivalent intensive longitudinal designs that included subjective self-report components.

**Exclusion criteria**

- Nonempirical publications, including reviews and conference abstracts.
- Studies based exclusively on objective physiological monitoring without subjective self-report components.

Two researchers independently screened titles and abstracts and then reviewed full texts for eligibility, with disagreements resolved through discussion or adjudication by a third senior researcher. The study selection process is shown in Figure 2. For each included study, we extracted bibliographic information, methodological characteristics from the Methods section (eg, EMA procedures, prompting design, tool or platform used, and analytic approach), and implementation-relevant descriptions and limitations from the Methods and Discussion sections. Extracted tool or platform information was used to descriptively summarize implementation approaches across studies.

We then conducted an iterative, workflow-based synthesis of the extracted implementation-relevant descriptions and limitations. Two researchers independently reviewed the extracted material and identified recurring implementation patterns and workflow-level gaps across studies. Through comparison and discussion, these recurring issues were organized into higher-order functional domains spanning the EMA study lifecycle. Discrepancies were resolved through discussion and reconciliation. The resulting domain structure was used to organize the synthesis of workflow patterns and implementation gaps across studies.

Detailed study-level evidence extraction and domain-linked gap synthesis are provided in Multimedia Appendix 1.

## MEPEF Development and Platform Benchmarking

### *Expert Panel and Framework Development*

We convened a purposively sampled multidisciplinary panel of 6 experts to support framework development from both mental health research and platform implementation perspectives. Panelists were recruited to provide disciplinary coverage across clinical psychiatry, psychotherapy, mental health research, interaction design, product management, and software engineering. Eligibility required at least 5 years of relevant professional experience in mental health research or mobile health development and familiarity with EMA study designs. This process was conducted as a moderated workshop-based consensus exercise and was intended to generate a preliminary implementation-oriented framework for subsequent testing and refinement.

Initial candidate indicators were generated from key workflow requirements identified in the landscape review. Using the 6 workflow-aligned domains from that review as an organizing structure, the panel refined these candidates through a moderated workshop-based consensus process. Across 2 iterative rounds of review and discussion, indicators were proposed, revised, merged, or removed as needed. Indicators were retained when panel members agreed that they were relevant to EMA research workflows, conceptually distinct from other indicators, and feasible for observable cross-platform assessment. The final indicator set was operationalized as the Multi-dimensional EMA Platform Evaluation Framework (MEPEF), a consensus-derived framework for cross-platform assessment.

### *Platform Benchmarking*

#### Platform Identification and Selection

The same multidisciplinary panel then used the MEPEF to benchmark 7 platforms selected to reflect distinct implementation models relevant to localized EMA deployment in China. Candidate platforms were identified from multiple complementary sources, including platforms reported in the landscape review, supplemental public searches, and researcher- or expert-recommended platforms. Eligibility for benchmarking required relevance to recurring EMA research workflows and sufficient accessibility for structured evaluation through platform operation, trial use, installation, and/or documentation review. Because access conditions and verification depth varied across platforms, the benchmarking was designed as an exploratory comparison of implementation-relevant capability patterns rather than as a definitive product ranking. Detailed platform categories, source types, eligibility considerations, and inclusion rationale are provided in Multimedia Appendix 2. The platforms were anonymized as Platforms A–G in the main text to keep the analysis focused on implementation-relevant capability patterns and workflow trade-offs.

### Benchmarking Procedures and Rating Process

Benchmarking was conducted through scenario-based walkthroughs of key implementation tasks, including configuring a study, setting prompt schedules and triggering rules, managing participants, monitoring survey delivery and completion, and exporting data. Walkthroughs were completed under moderator guidance to ensure that experts covered the same core steps across platforms, while still allowing them to operate the platforms directly and consult available documentation as needed.

After completing each walkthrough, experts independently rated all applicable indicators based on hands-on use and document review. Indicators were scored on a 3-point scale (1=does not meet, 2=partially meets, 3=fully meets). Indicators initially judged as discrepant or uncertain were discussed to clarify platform functions and scoring criteria, after which experts could further inspect the platform and independently finalize their ratings. Indicators that could not be verified through platform operation or documentation review were left unrated.

### Cross-Platform Synthesis

For cross-platform synthesis, finalized independent indicator-level ratings were aggregated within the 6 MEPEF domains to generate domain-level mean scores for each platform. All benchmarking and interrater reliability analyses reported in the main text were conducted using the finalized MEPEF indicators. For visualization, these domain means were rounded to the nearest integer to produce a simplified 3-level profile matrix. To assess the consistency of expert ratings across platforms, interrater reliability was estimated from the finalized independent ratings using intraclass correlation coefficients based on a 2-way mixed-effects, absolute-agreement model, with average-measures coefficients treated as primary.

In addition to domain-level scoring, walkthrough observations and issues raised during post-walkthrough discussion were reviewed descriptively across platforms to identify exploratory cross-platform contrasts and recurrent implementation trade-offs within the 6 MEPEF domains. These observations informed the generation of benchmarking-derived implications for subsequent synthesis.

## Framework-Guided Design Targets and Case Mapping

Huixin EMAI was an existing localized platform before the present study and was examined here as a framework-guided case. Domain-specific implementation implications identified through the landscape review and platform benchmarking were synthesized within the 6 MEPEF domains and further consolidated into framework-guided design targets spanning protocol control, deployment workflow, multimodal readiness, data support, privacy-preserving local operation, and participant accessibility. The framework was then applied to Huixin EMAI to systematically examine current implementation coverage, with the derived design targets used to map relevant platform functions and inform priorities for refinement and adaptation. For each domain, this case mapping was conducted through direct walkthrough of platform functions, inspection of operational features, and review of representative interface components. The mapping was descriptive and intended to document current representation of framework-derived design targets in Huixin EMAI rather than to establish formal achievement or independent validation of each target.

Huixin EMAI (ZenSeven Technology Co., Ltd, Hangzhou, China) comprises 2 coordinated components: Huixin Assessment, a participant-facing mobile application, and Huixin Research, a researcher-facing web portal, both connected to a centralized backend. As shown in Figure 1, the backend supports study configuration, secure data ingestion and storage, scheduled and event-contingent prompting, and unified timestamping of EMA events, with wearable and multimodal modules included to support extensibility. The correspondences between framework-derived design targets and operational features of Huixin EMAI are illustrated through representative interface components and system functions.

**Figure 1.** System architecture of the Huixin EMAI platform. Solid arrows indicate data flow and dashed arrows indicate notification flow. Wearable and multimodal modules are shown to illustrate extensibility.

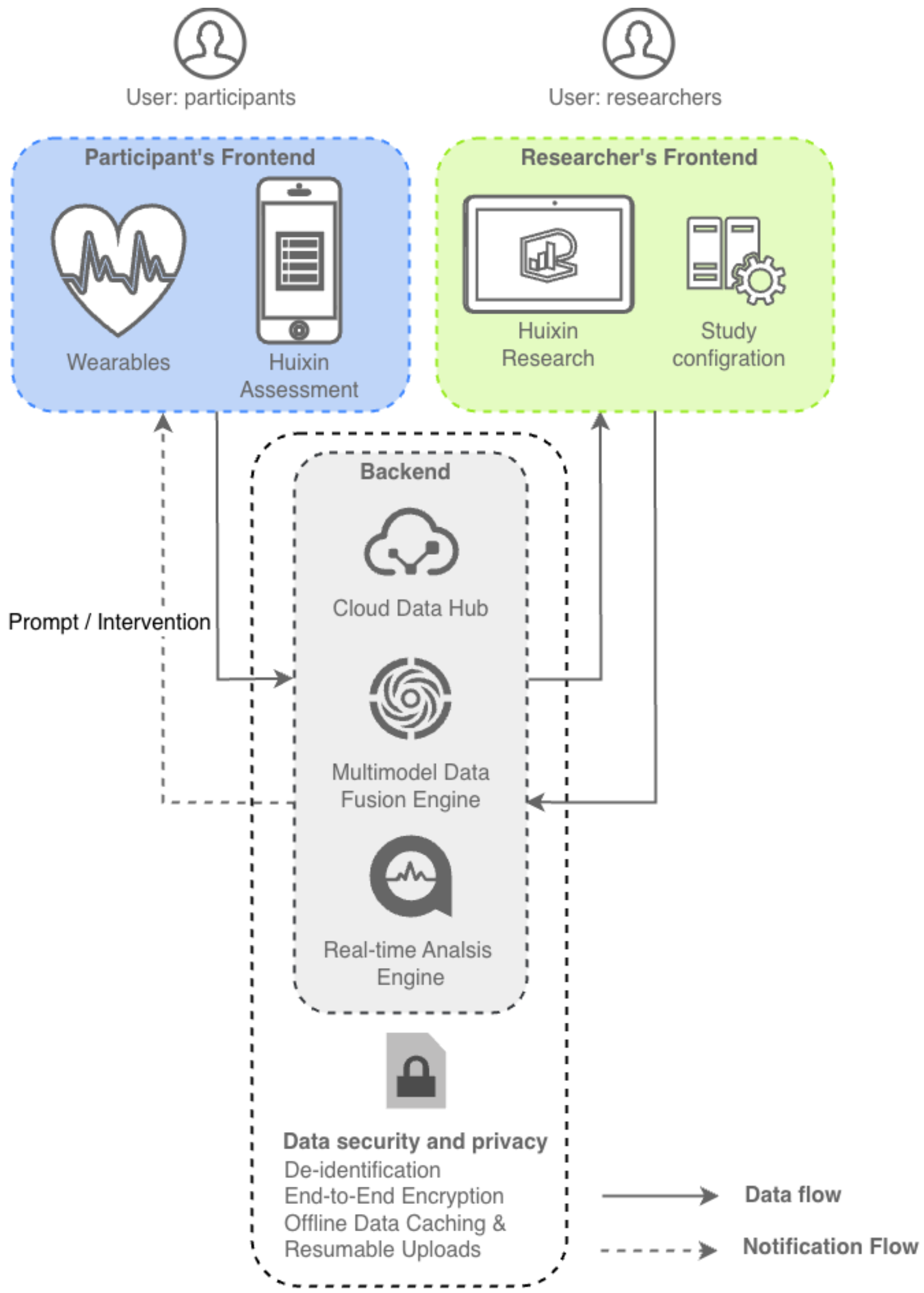

## Preliminary Implementation Evaluation of the Huixin EMAI

### *Case Study Deployment and Quantitative Implementation Indicators*

Huixin Assessment was deployed in an EMA study involving 48 adult participants, including 26 healthy participants and 22 patients with anxiety disorders. All participants used their own devices to access the app, including 27 Android devices and 21 iOS devices. The overall data collection window spanned approximately 2 weeks because of study scheduling logistics, whereas the EMA monitoring period for each participant was approximately 8 study days.

During monitoring, 6 randomly scheduled prompted questionnaires per day were delivered between 09:00 and 21:00. Participants were encouraged to complete these prompted questionnaires as promptly as possible, and each prompted questionnaire expired 30 minutes after delivery if unanswered. In addition, resident questionnaires remained available for voluntary self-initiated completion. Because prompted and resident questionnaires served different purposes, response latency was calculated only for valid prompted responses and was therefore interpreted as an indicator of timely responding rather than overall completion behavior across all EMA entries. In parallel with subjective data collection, participants also wore a third-party wearable device, the Hexoskin ProShirt (Carré Technologies Inc., Montréal, Canada), to collect respiratory physiological data. These physiological data were subsequently matched to EMA records by timestamp, and subjective-objective temporal linkage was defined as the proportion of EMA records with breathing rate, tidal volume, and minute ventilation all valid within 5 minutes. Quantitative implementation indicators were calculated from backend records and included study days, completed EMAs per study day, prompt completion rate, response latency for valid prompted responses, and subjective-objective temporal linkage.

### *Participant Feedback on Huixin Assessment*

Subjective feedback on Huixin Assessment was collected during a scheduled telephone debriefing conducted immediately after completion of the EMA study to minimize recall bias. During this follow-up, research staff administered a study-specific structured questionnaire intended for acceptability assessment that included 5-point Likert items (1 = strongly disagree, 5 = strongly agree) and a multiple-choice item on barriers to timely completion. The questionnaire was informed by constructs from the System Usability Scale and prior EMA acceptability instruments [16,17]. It included 9 Likert-type items assessing comprehensibility, usability, perceived burden, privacy, and willingness to sustain participation, as well as structured follow-up questions on reasons for missed or delayed responses and optional open-ended feedback. Questionnaire responses were summarized descriptively and were not intended to provide formal psychometric validation.

### *Researcher Interview and Qualitative Analysis*

We also collected qualitative input from 6 researchers who had used Huixin Research in independent studies for at least 3 months. Three had greater involvement in earlier study planning and protocol design, whereas the other 3 were primarily responsible for day-to-day platform operation during study implementation. Using a semi-structured interview guide organized around the overall EMA workflow, YL conducted individual interviews on experiences with the researcher-facing component during real-world study implementation. Interviews lasted approximately 20 to 40 minutes and were documented with permission in deidentified interview notes. The interview guide is provided in Multimedia Appendix 3.

Interview data were analyzed using directed qualitative content analysis guided by the 6 MEPEF domains as an a priori framework. Two coders independently applied a shared codebook. The first coding layer assigned each segment to 1 domain, and the second classified it as either an achieved feature or a remaining limitation. Coding discrepancies were resolved through reconciliation, and intercoder reliability was assessed using Cohen kappa. A total of 372 relevant coding units were included and 185 nonrelevant units were excluded. Percent agreement was 84.95% (κ=0.806) for domain coding and 69.62% (κ=0.636) for combined domain-plus-status coding. Only deidentified interview data are reported in this study.

### Ethics Statement

Participant data were collected within an EMA study approved by the Ethics Committee of Hangzhou Seventh People's Hospital (No. 2024/037). Researcher interviews were conducted separately from the participant EMA study as stakeholder consultation for platform improvement using deidentified note-based documentation. All study participants and researcher interviewees were informed about the purpose of the study and the intended use of their data and provided informed consent before participation.

## Results

### Landscape Review Findings and Workflow-Level Gaps

As shown in Figure 2, the database search yielded 241 records (CNKI n=151; Wanfang Data n=90). After removing 32 duplicates, 209 unique records were screened, and 66 empirical studies were included. These studies were used to characterize implementation approaches in Chinese mental health EMA research and to identify recurring workflow-level gaps relevant to localized deployment.

**Figure 2.** Flow diagram of the literature screening and selection process.

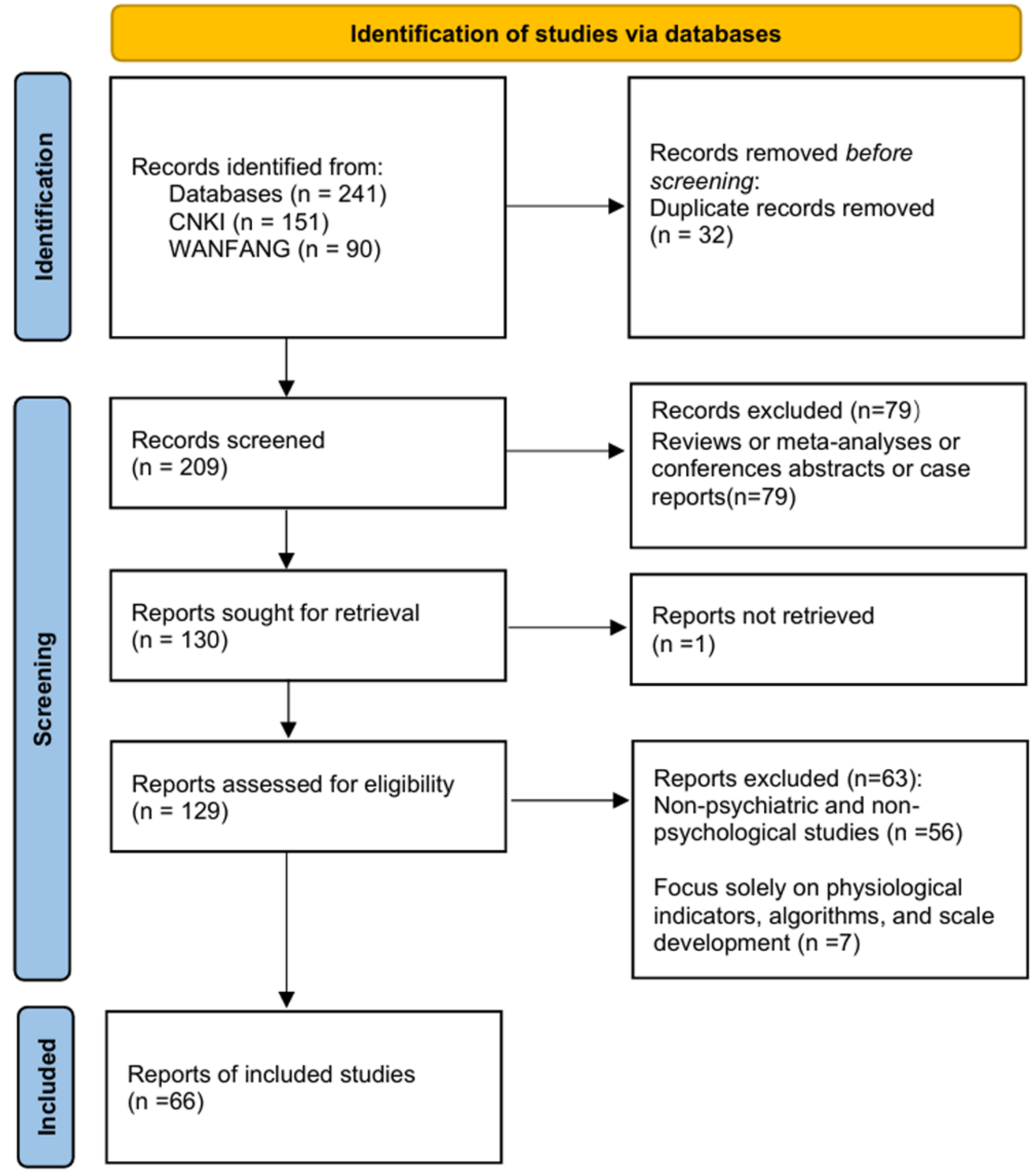

Based on extracted tool and platform information, implementation approaches were summarized into 4 categories. The most common approach was instant-messaging–based workflows (36/66, 54.5%) [18–53], usually combined with general survey platforms, followed by specialized EMA platforms (14/66, 21.2%) [54–67] and paper-based diaries (11/66, 16.7%) [68–78]. In 5 of 66 studies (7.6%), the implementation tool or platform was not reported, limiting reproducibility and cross-study comparability [79–83].

Descriptively, publication volume was higher in later years, with 18 studies published before 2021 and 48 published from 2021 onward. Specialized EMA platforms were reported in 2 of 18 studies (11.1%) before 2021 and 12 of 48 studies (25.0%) from 2021 onward, although instant-messaging–based workflows remained the most common implementation approach overall (36/66, 54.5%) and in both periods (10/18, 55.6% before 2021; 26/48, 54.2% from 2021 onward).

The thematic synthesis identified 6 workflow-aligned domains spanning the EMA study lifecycle. As summarized in Table 1, these domains captured recurring workflow-level gaps reported across included studies and were used to organize representative gap statements and supporting evidence.

**Table 1.** Workflow-aligned domains, synthesized gap statements, and supporting studies identified in the review of Chinese mental health EMA studies

| Domain | Major synthesized gap statement | Supporting studies |
|---|---|---|
| Research design and configuration | Fixed-time sampling predominated, with limited use of randomized or event-contingent schedules. | [19,23–26,28,30,32,36,40–43,48–51,60,71,75] |
| | Sampling intensity or study duration was sometimes insufficient to capture dynamic symptom variation. | [25,26,32,33,35,39,46,52,55,56,58–61,64,65,68,74,76,80,83] |
| Implementation and deployment | Prompt delivery, completion tracking, incentive administration, and follow-up often relied on manual researcher effort. | [18–53,55,57,68–79] |
| | Participant onboarding, training, and pre-study practice imposed substantial implementation burden. | [19,20,26,28–30,32–34,36,37,39,40,44,49,52,53,55,56,58,59,62,65,67,68,71,72,75–79,81] |
| Extension and integration | EMA workflows remained largely self-report based, with limited wearable or mobile-sensing integration. | [19,21,26,27,32,39,41,43,45,50,55,57,60,65,66,80] |
| | Integration of objective data remained limited, and in some studies such data still depended on screenshot submission and manual transcription. | [32,36,63,73] |
| Data management | Real-time compliance support was limited. | [19,20,22,26,28,29,32,36,39,40,44,49,53,55,57,68,79] |
| | Built-in support for data processing and visualization was limited. | [18–24,26,27,29,30,32,33,35–39,41,42,44–49,51–55,59,61,65,66,68,72–75,77–79,81–83] |
| Ethics, privacy, and security | Account-linked coordination and traceable coding practices created privacy and re-identification risks. | [20,22,23,25–32,35–41,43–45,47,49,55,65] |
| Accessibility | Specialized EMA software was associated with accessibility barriers for older adults, lower-literacy users, or adolescents with restricted device access. | [54,63,64] |
| | Successful deployment often depended on substantial onboarding training and responsive technical support during the study period. | [34,55,56,58,59,65,67] |

Note: Table 1 presents the major synthesized gap statements identified across studies. More detailed study-level evidence mapping and additional domain-linked gap statements are provided in Multimedia Appendix 1.

### Development of MEPEF

The expert consensus process resulted in the Multi-dimensional EMA Platform Evaluation Framework (MEPEF), comprising 43 observable indicators across 6 domains. These indicators were refined from an initial set of 26 workflow-oriented indicators through expert review, including item splitting, consolidation, and wording refinement to improve observability and cross-platform applicability. The 6-domain structure identified in the landscape review was retained, with minor label refinements for framework application. The full indicator set is presented in Textbox 2, and detailed operational definitions and rating guidance are provided in Multimedia Appendix 4.

**Textbox 2.** Multi-dimensional EMA Platform Evaluation Framework (MEPEF)

1. Study design and configuration

- Survey item diversity
- Survey logic and flow control

- Sampling design and triggering configuration
- Prompt scheduling and reminder configuration
- Response-window and late-response handling
- Group allocation and protocol differentiation
- Protocol versioning and change history
- Pre-deployment testing and simulation

2. Implementation and deployment

- Electronic informed consent
- Participant onboarding and enrollment management
- Prompt delivery reliability and permission handling
- Notification content and delivery settings
- Compliance monitoring and follow-up reminders
- Offline response capture and synchronization
- Response editing and resubmission handling
- Application stability and operational reliability
- Researcher-side deployment control

3. Extension and integration

- Wearable and sensor data synchronization
- Third-party API and external integration support
- Support for multimodal data linkage
- External data import and record matching

4. Data management

- Event-level timestamping capture
- Process metadata capture and export
- Export completeness and field clarity
- Analysis-ready export
- Selective export by time range or subgroup

- In-platform monitoring summaries
- Basic descriptive statistics and visualization
- Data-quality checking and anomaly flagging

5. Ethics, privacy, and data security

- Consent documentation and access auditability
- De-identification support
- Encryption in transit and at rest
- Access control and role-based permissions
- Data sharing and export governance
- Data deletion and account cancellation support
- Privacy-preserving participant management

6. Usability and accessibility

- Cross-platform client support
- Public availability and trial access
- Multi-language support
- Researcher-side and participant-side usability
- User feedback channels for issue reporting during studies
- Training and support materials
- Accessibility support for diverse user groups

### MEPEF-Based Platform Benchmarking

MEPEF-based benchmarking showed clear cross-platform differences in domain-level capability profiles across the 6 domains (Figure 3). Platforms E–G generally showed higher and more balanced profiles across domains, whereas Platforms A–D more often showed uneven capability patterns, particularly in extension and integration and implementation-related domains.

**Figure 3.** Cross-platform capability benchmarking matrix across 6 MEPEF domains.

| Platform | Study design and configuration | Implementation and deployment | Extension and integration | Data management | Ethics, privacy, and security | Usability and accessibility |
|---|---|---|---|---|---|---|
| Platform A | | | | | | |
| Platform B | | | | | | |
| Platform C | | † | † | | | |
| Platform D | | † | | | | |
| Platform E | | † | | | | |
| Platform F | | | | | | |
| Platform G | | | | | | |

Legend
1 = does not meet
2 = partially meets
3 = fully meets
† = added when at least 2 indicators within that domain were marked as unverified by 3 or more raters

At the indicator level, unrated entries were primarily attributable to functions that could not be verified through platform operation or documentation review. Of 1806 indicator-level ratings, 113 (6.3%) were left unrated. After aggregation to platform-domain-rater scores, all 252 scores could be derived. Dagger symbols in Figure 3 identify platform-domain cells in which unverified ratings were relatively concentrated. Domain-level mean scores and SDs across raters for each platform are provided in Multimedia Appendix 2.

Domain-level interrater reliability was high overall. Using the 43-indicator scoring structure, intraclass correlation coefficients (ICCs) were calculated across all 7 platforms in each domain. Average-measures ICCs ranged from 0.78 to 0.97 across domains, indicating that domain scores aggregated across raters were relatively stable. Domain-level average-measures ICCs were 0.92 for study design and configuration, 0.92 for implementation and deployment, 0.87 for extension and integration, 0.97 for data management, 0.78 for ethics, privacy, and data security, and 0.91 for usability and accessibility.

Beyond the domain-level profiles shown in Figure 3, expert walkthroughs and subsequent discussion identified several implementation-level contrasts that were not fully captured by domain scores alone. These observations are summarized in Table 2.

**Table 2.** Observations from expert benchmarking

| Domain | Benchmarking observations |
|---|---|
| Research design and configuration | Platform B-D showed more limited support for flexible study configurability, including branching logic and longitudinal scheduling options. |
| Implementation and deployment | Platform A was relatively easy to incorporate into familiar WeChat-based workflows, but routine deployment still depended heavily on manual coordination. |
| | Platforms E–G were generally usable under benchmarking conditions, but appeared less suitable for routine participant-facing deployment in local settings. |
| Extension and integration | Platform A provided limited integrated multimodal support, whereas Platforms B and C often handled wearable-related components outside the survey workflow. |
| Data management | Platforms B–D generally supported basic study monitoring and post-study data export but offered more limited support for real-time management and analysis-ready data handling. |
| Ethics, privacy, and security | Platforms E and F showed recurrent implementation frictions related to nonlocal data hosting and misalignment with local health-data governance requirements. |
| Usability and accessibility | Platform B showed inconsistent notification delivery on some Android devices, whereas Platform D did not support iOS users. |
| | Platforms E–G faced barriers to routine local use, including app distribution and maintenance challenges, as well as compatibility issues across local networks, devices, and operating systems. |

### Framework-Derived Design Targets and Mapping to the Huixin EMAI Case

The landscape review and cross-platform benchmarking together pointed to 6 domain-specific design targets for localized EMA implementation: research-grade protocol control, lower-burden implementation, multimodal integration readiness, analysis-ready data support, privacy-preserving local deployment, and participant accessibility. Table 3 summarizes these targets and their current representation in Huixin EMAI.

Overall, the mapping suggested that Huixin EMAI showed observable alignment with the major framework-guided design targets across the 6 MEPEF domains, although the extent and maturity of current representation varied across domains. Figures 4 and 5 provide illustrative examples from the researcher-facing and participant-facing components of the platform, respectively.

**Table 3.** Framework-derived design targets and their current representation in the Huixin EMAI case

| Domain | Landscape review-derived implications | Benchmarking-derived implications | Framework-derived design targets | Current representation in Huixin EMAI |
|---|---|---|---|---|
| Research design and configuration | Support fixed, randomized, and event-contingent scheduling with configurable prompt rules; enable flexible sampling intensity and study duration to better capture symptom dynamics. | Configurable branching, flexible scheduling, and group-specific protocol configuration for research-grade EMA. | Research-grade protocol control | Multiple prompting logics, configurable response rules, and differentiated protocol assignment for research-grade EMA protocols (Figure 4, panels 1–4). |
| Implementation and deployment | Automate prompt delivery, completion tracking, reminders, and incentive workflows; streamline onboarding with guided enrollment, training, and test-mode practice. | Automated prompting, reminders, compliance tracking, and deployment control; lower-burden onboarding, participant support, and localization-ready deployment. | Lower-burden implementation | Automated prompting and reminder workflows, participant onboarding and support, routine monitoring, and researcher-side deployment oversight (Figure 4, panels 5–6; Figure 5, panels 1–3). |
| Extension and integration | Support multimodal integration with wearables, sensors, and external data sources; enable direct objective-data capture, | Support both native and externally linked integration of objective physiological, wearable, and environmental data. | Multimodal integration readiness | Device binding, data transmission, and timestamp-based alignment for EMA-linked wearable and sensor data |
| | timestamp alignment, and structured export. | | | (Figure 4, panels 7–8; Figure 5, panels 5–6). |
| Data management | Provide real-time compliance monitoring at participant and study levels; provide analysis-ready exports and basic summaries and visualizations. | Provide real-time monitoring, process-level records, and analysis-ready export. | Analysis-ready data support | Operational monitoring, execution records, and filterable exports for downstream analysis and quality checking (Figure 4, panels 6 and 8–9). |
| Ethics, privacy, and security | Use de-identified participant management and minimize identifiable data exposure; strengthen security through encrypted transmission, storage, and role-based access control. | Enable local data hosting, compliant storage, de-identification, encryption, and permission-controlled access. | Privacy-preserving local deployment | Local hosting, de-identified participant management, encrypted data handling, and permission-controlled access (Figure 4, panel 5; Figure 5, panel 6). |
| Usability and accessibility | Improve accessibility through simple interfaces and low-burden interaction design; include in-app guidance and support tools to reduce | Reliable cross-platform notification delivery and full iOS-Android support; localization-ready app distribution and robust local | Participant accessibility | Simple participant workflows, in-app guidance, routine app access, and cross-platform compatibility (Figure 5, |
| | technical barriers. | device compatibility. | | panels 1–4 and 6). |

**Figure 4.** Illustrative researcher-facing interfaces from Huixin Research corresponding to framework-derived design targets.

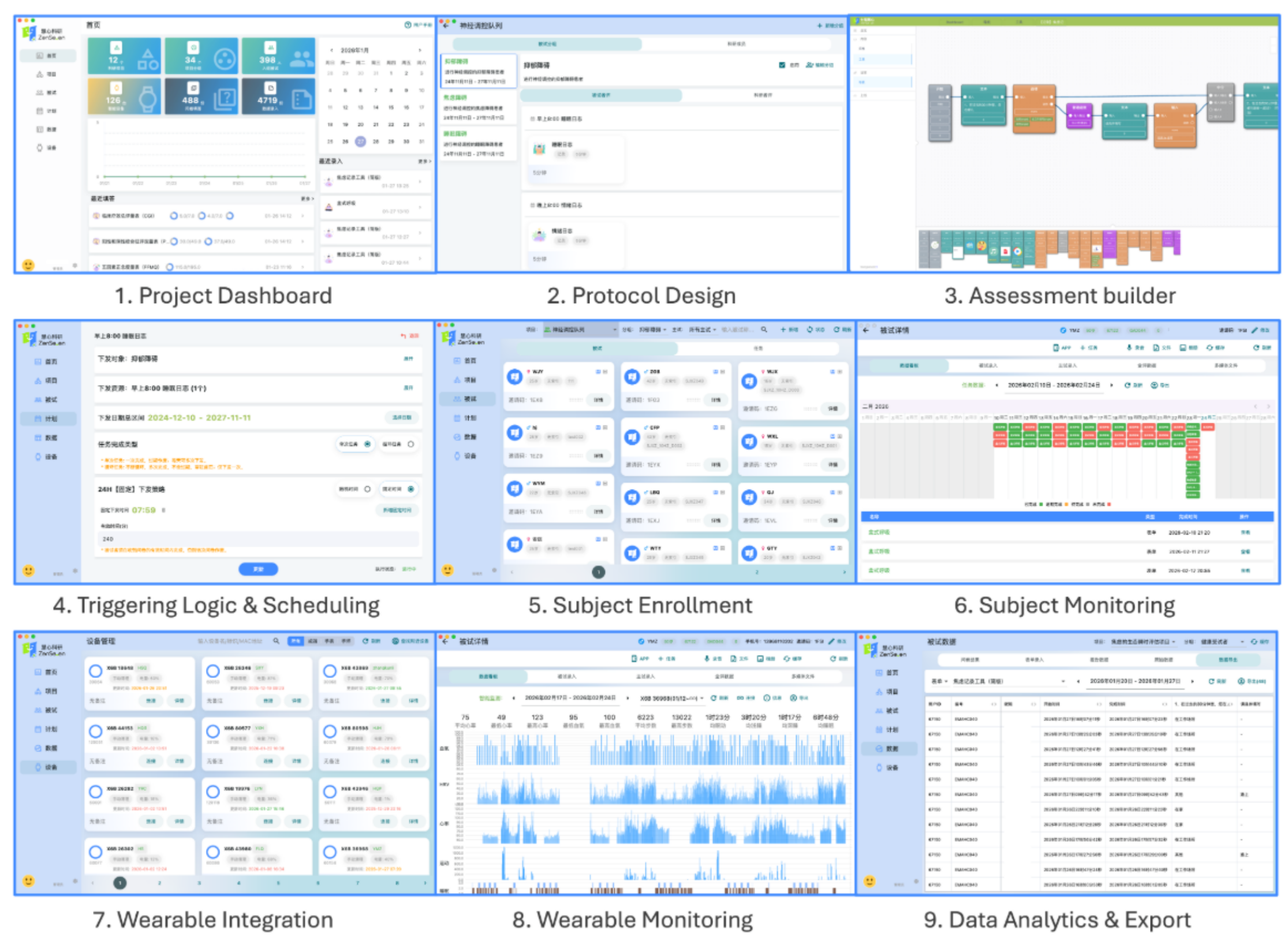

**Figure 5.** Illustrative participant-facing interfaces from Huixin Assessment corresponding to selected framework-derived design targets.

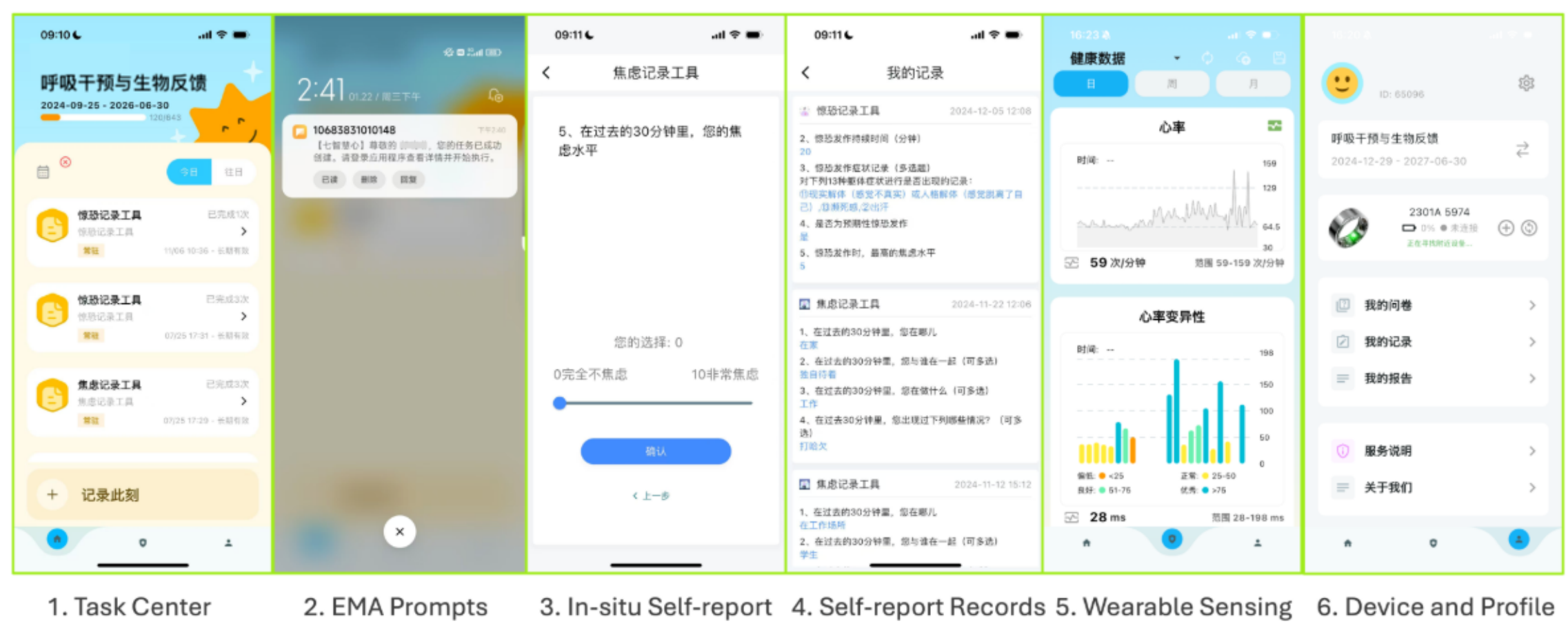

## Deployment Performance and User Feedback

### *Objective Deployment Performance*

During the deployment period, 48 participants contributed EMA data, including 26 healthy participants and 22 participants with anxiety disorders. Overall, 1893 of 2472 expected prompts were completed, corresponding to a prompt completion rate of 76.6%. Participants contributed data across a mean of 8.58 study days (SD 2.48) and completed a mean of 4.52 EMAs per study day (SD 1.32). Among valid prompted responses, the response-level median latency was 4.00 minutes (Q1–Q3 0.00–13.00), and subjective-objective temporal linkage reached 82.1%.

Of the completed EMA entries, 1122 valid prompted responses were included in latency analysis, whereas 771 entries were resident questionnaires that were not latency-eligible. In exploratory subgroup comparisons, the healthy group showed a higher prompt completion rate (82.0% vs 69.8%; P=.018) and more completed EMAs per study day (4.94 vs 4.03; P=.018). Android and iOS users showed similar prompt completion rates and EMA completion levels, but Android users showed longer mean latency for valid prompted responses than iOS users at the subject level (11.32 vs 5.58 minutes; P<.001). Detailed overall and subgroup results are presented in Multimedia Appendix 5. Because these subgroup analyses were exploratory, no adjustment for multiplicity was performed, and the associated P values are presented for descriptive interpretation.

### *Participant-Reported Acceptability and Barriers to Timely Completion*

Participant ratings indicated generally favorable acceptability of Huixin Assessment, particularly for comprehensibility, ability to use the app normally, sustained completion across multiple days, ability to complete 4 or more questionnaires per day, and perceived privacy (Table 4). Ratings for negatively worded items were comparatively low, suggesting that notification burden and interruption to daily activities were present but generally acceptable.

**Table 4.** Participant-Reported Acceptability Ratings for Huixin Assessment (N=44)

| Item | Mean (SD) |
| --- | --- |
| I understood how the Huixin Assessment app works. | 4.61 (0.69) |
| During the study, I could use the Huixin Assessment app normally on my phone. | 4.36 (0.84) |
| Completing questionnaires on the Huixin Assessment app using my smartphone was easy. | 4.34 (1.06) |
| The questionnaires in the Huixin Assessment app were displayed well on my smartphone. | 4.11 (1.10) |
| During the study, I completed the app-prompted questionnaires for multiple consecutive days. | 4.64 (0.61) |
| During the study, I completed 4 or more questionnaires per day in the Huixin Assessment app. | 4.55 (0.79) |

| Item | Mean (SD) |
|---|---|
| The number of app notifications was bothersome.* | 2.68 (1.20) |
| Completing the app-prompted questionnaires interrupted my daily activities.* | 2.86 (1.07) |
| I could complete the daily questionnaires privately without worrying that others would see my responses. | 4.59 (0.84) |

Note: All items were rated on a 5-point Likert scale. Items marked with an asterisk are negatively worded; lower scores indicate more favorable experiences.

Barriers to timely completion were mainly contextual and notification-related. The most frequently endorsed reasons were being too busy (31/44, 70.5%) and not noticing the notification in time (31/44, 70.5%), followed by being asleep (14/44, 31.8%). Few participants attributed delayed completion to not having their phone with them (6/44, 13.6%), survey burden (2/44, 4.5%), privacy concerns (1/44, 2.3%), or lack of motivation (1/44, 2.3%).

### *Researcher Feedback on Current Platform Strengths and Remaining Frictions*

#### Study Design and Configuration

Researchers described Huixin Research as supporting core EMA design functions, including relatively complex questionnaire logic, fixed and random prompting, repeated measurement, and some degree of protocol differentiation. As R03 noted, researchers could specify whether prompts were sent "on a fixed schedule or randomly," along with start and end times, response windows, and "some personalized prompting." However, these functions still depended substantially on developer or backend support rather than full researcher-side control. As R06 noted, "most questionnaire or task settings still require contacting backend staff and waiting for review before going live."

#### Implementation and Deployment

Researchers described the platform as already supporting core implementation workflows, including onboarding, questionnaire delivery, automated prompting, and real-time completion monitoring. As R06 stated, the system "can send [questionnaires] automatically," which "greatly reduces some of the manual effort." However, weaknesses remained in proactive delivery alerts, adherence reminders, and researcher-side process control. As R05 noted, "if a notification was not delivered, we need to be reminded so that we can contact the participant."

#### Extension and Integration

Researchers viewed the platform as having a workable baseline for multimodal EMA through wearable linkage and researcher-side physiological data display. As R06 explained, the wearable connects to the participant's phone through Bluetooth, after which data can also be viewed on the research side. However, this integration remained limited by connection problems. As R01 reported, patients sometimes "keep saying the hardware gets disconnected."

### Data Management

Researchers described the platform as supporting basic data viewing, time-range selection, and batch export. As R01 noted, researchers could “select a time period and download multiple participants.” However, downstream usability remained a prominent weakness, particularly because exported data still required additional cleaning and some summary indicators had to be recalculated manually. As R05 reported, “the exported format and item positions” could change across application versions.

### Ethics, Privacy, and Data Security

Researchers reported that Huixin Research included basic privacy-protection and role-differentiation mechanisms, including deidentification, masked phone numbers, and account-based permission assignment. As R04 noted, “the data we receive are anonymized,” and only authorized researchers can view participant information. However, governance flexibility and permission clarity remained limited, sometimes restricting routine operations for non-administrator accounts. As R05 explained, some permissions “are not very clear.”

### Usability and Accessibility

Researchers generally viewed the interface as understandable and manageable for both researchers and participants. However, they also reported routine usability frictions, including slow loading, occasional unresponsiveness, and relatively plain completion interfaces with limited validation support. As R06 noted, “its interface is relatively simple and easy to use,” whereas R01 described the completion interface as feeling “more like filling in a text box.”

Taken together, the interviews suggested that Huixin Research could support routine EMA deployment in practice, while also highlighting remaining frictions in researcher-side configurability, delivery monitoring, multimodal stability, downstream data usability, governance support, and interface smoothness.

## Discussion

### Principal Findings

The main barrier to advancing EMA in Chinese mental health research appears to lie less in basic feasibility itself than in the lack of implementation pathways that are locally deployable, sustainable, and suitable for routine use. The landscape review showed that EMA is already being conducted in China, but its implementation still relies largely on labor-intensive and relatively inflexible workflows that are difficult to sustain and scale across studies. Benchmarking revealed a persistent trade-off: platforms with stronger research-oriented functionality were often less compatible with local deployment conditions, whereas more readily deployable approaches tended to offer weaker protocol control, automation, and integration capacity.

Together, these findings suggest a structural mismatch between the requirements of research-grade EMA and the realities of implementation in the Chinese context.

To address this gap, we developed MEPEF as an implementation-oriented evaluation framework that translated workflow-level requirements into assessable domains and framework-guided design targets. Informed by the landscape review and refined through expert workshops, MEPEF structured platform assessment around key EMA workflow domains and clarified core requirements for localized EMA platforms. The resulting design targets were research-grade protocol control, lower-burden implementation, multimodal integration readiness, analysis-ready data support, privacy-preserving local deployment, and participant accessibility. When applied to the Huixin EMAI case, the framework helped assess current implementation coverage and interpret how existing platform functions aligned with these design targets. The preliminary deployment findings suggest that the platform provides preliminary support for core localized EMA workflows, while still revealing important gaps in researcher-side control, delivery monitoring, data readiness, and governance support. These findings should therefore be interpreted as evidence of preliminary implementation coverage rather than as proof of platform effectiveness.

### Contribution and relation to prior work

Prior work has shown that many research-grade EMA capabilities are technically feasible and that a broad ecosystem of EMA tools already exists. Existing studies have provided platform selection guidance, mapped the EMA technology design space, cataloged available tools, and proposed multiple ways to lower technical barriers, including no-code protocol authoring, monitoring-oriented research platforms, and open-source solutions that support self-hosting and data control [7,12,13,84–97]. These studies have been highly informative for understanding what EMA platforms can offer and how different technical choices may support different study goals.

However, most of this literature is oriented toward platform capability, design choice, or general tool availability rather than toward the practical problem of localized implementation. In particular, it offers limited guidance on how to move from technically available EMA functions to workflows that are practically deployable, operationally supportable, and reusable under specific local constraints. This limitation is especially relevant in settings such as China, where platform accessibility, domestic device compatibility, data governance requirements, and long-term operational support may shape feasibility as much as technical functionality itself. Even when China-based implementations have been reported, they have often appeared either as study-specific solutions that are difficult to generalize across projects or as highly accessible instant-messaging–based workflows, often combined with general survey platforms, that lower entry barriers but provide limited support for automation, monitoring, and protocol governance in more complex EMA designs [98–100].

Our study adds to this literature by moving beyond feature comparison to examine implementation pathways. Rather than asking only what EMA platforms can do, we examined how workflow demands emerging from published mental health EMA studies can be translated into observable evaluation domains, cross-platform capability trade-offs, and framework-guided design targets for localized EMA platforms. The contribution of this study lies not in de novo platform development, but in demonstrating how a framework grounded in local workflow realities can be used to systematically assess an existing localized platform, clarify its current implementation coverage, and identify remaining priorities for refinement. Although this work was conducted in the Chinese mental health context, the underlying contribution is not limited to China. The framework and pathway presented here may also be useful in other settings where EMA implementation is shaped by local deployment frictions, heterogeneous device ecosystems, governance requirements, and the need to balance research-grade functionality with operational feasibility. In this sense, China functions here not only as an application context but also as a high-constraint implementation setting in which broader infrastructure challenges become especially visible. Nevertheless, MEPEF should still be regarded as a preliminary implementation-oriented framework that warrants further external validation across settings.

## Limitations

This study has several limitations. First, the landscape review was restricted to published studies in two major Chinese databases to focus on EMA practices under local deployment conditions and may not have captured unpublished, commercial, institution-specific, or internationally developed EMA implementations not represented in this literature. Second, the preliminary evaluation of the Huixin EMAI case involved limited participant and researcher samples and a relatively short implementation period, limiting the generalizability of the findings. Researcher feedback may also have been influenced by positive user experience bias, as all interviewees had experience using the platform. The qualitative analysis was based on deidentified interview notes rather than verbatim transcripts. Third, benchmarking verification depth varied across platforms because access conditions, trial functions, and documentation completeness were not identical, and some platform functions remained unverified rather than fully testable. Fourth, because MEPEF was developed and applied within a connected research process, the present findings primarily support its preliminary organizing utility for implementation-relevant assessment and refinement priorities within this study context, rather than completed external validation. Further application across independent settings will be important for examining its broader generalizability and robustness.

## Conclusion

The advancement of EMA in mental health research depends not only on the availability of technically capable tools, but also on the ability to embed those tools within workflows that are operationally workable and sustainable over time. Through a landscape review, cross-platform benchmarking, and application to an existing localized case, this study showed how workflow-level implementation gaps can be translated into a structured evaluation framework, framework-guided design targets, and a practical case-based assessment pathway. MEPEF provided an implementation-oriented framework for this process, and the Huixin EMAI case illustrated how current implementation coverage can be assessed and remaining priorities for refinement identified in a localized platform context. Overall, these findings may inform future efforts to advance localized EMA platforms under local constraints in China and in other settings facing similar deployment challenges.

## Acknowledgments

During manuscript preparation, the authors used DeepSeek-V3.2 to assist with spelling and grammar checks. The authors reviewed and edited all AI-assisted output and take full responsibility for the content of the manuscript.

## Funding Statement

This work was supported by the Construction Fund of Key Medical Disciplines of Hangzhou (grant number: 2025HZZD14).

## Data Availability

Huixin EMAI includes the participant-facing app “慧心评估 (Huixin Assessment),” available on Android and iOS, and the researcher-facing system “慧心科研 (Huixin Research),” available on Android, Windows, and macOS. Access for research use can be requested from the corresponding author. Deidentified research materials supporting the present evaluation may be made available on reasonable request, subject to ethics requirements, participant privacy protection, and commercial constraints related to the platform.

## Conflicts of Interest

RG is a product manager at ZenSeven Technology Co., Ltd, and CHY is the general manager of ZenSeven Technology Co., Ltd. ZenSeven Technology Co., Ltd contributed to the development of the Huixin EMAI platform described in this study. XZ and JW were involved in early requirement discussions related to the platform. The roles of RG and CHY were limited to platform development and implementation support. The academic research team independently designed the study, conducted the evaluation, performed the analyses, and interpreted the findings, and finalized the manuscript text. The other authors declare no competing interests.

## Authors' Contributions

XZ, YL, and JW contributed equally to this work and share first authorship. XZ and CY conceived the study. XZ, YL, JW, CY, and WD designed the methodology. XZ, YL, and YF conducted the literature review, screening, data extraction, and evidence synthesis. XZ, YL, JW, and YF conducted the field evaluation, qualitative data collection, and organization of study materials. XZ, YL, and JW developed the evaluation framework and conducted platform benchmarking. JW performed the formal analysis. XZ drafted the original manuscript. CY and WD supervised the study. RG and CHY contributed to the technical development and implementation support of Huixin EMAI. All authors reviewed, revised, and approved the final manuscript.

## Abbreviations

API: application programming interface CNKI: China National Knowledge Infrastructure EMA: ecological momentary assessment ESM: experience sampling method ICC: intraclass correlation coefficient JITAI: just-in-time adaptive intervention MEPEF: multi-dimensional EMA platform evaluation framework mHealth: mobile health SDK: software development kit